\documentstyle[twocolumn,eqsecnum,aps,epsf]{revtex}

\def\JH{J_{\rm H}}
\def\Ham{{\cal H}}
\def\Re{{\cal R}\!{\sl e}\,}
\def\Im{{\cal I}\!{\sl m}\,}

\def\HSIZE{8cm}

\begin{document}
\title{Scaling relation of spin wave lifetime in double-exchange 
systems}
\author{Nobuo Furukawa\cite{email}}
\address{
  Institute for Solid State Physics,  University of Tokyo,\\
 Roppongi 7-22-1,  Minato-ku, Tokyo 106, Japan
}
\author{K. Hirota}

\address{
Department of Physics, Tohoku University,\\
  Aramaki Aoba, Sendai 980-77, Japan
}

\date{\today}
\maketitle

\begin{abstract}
Spin excitation spectrum of CMR manganites are studied
from both theoretical and experimental sides.
Scaling relations for the spin wave lifetime in the
 electronic model with double-exchange interaction is observed,
which is consistent with results for neutron inelastic scattering
experiments for (La,Y)$_{0.8}$Sr$_{0.2}$MnO$_3$.
Roles of other interactions and degrees of freedom such as
polaron effects are also discussed.
\end{abstract}

\vspace{2mm}
To be published in Physica B.


\section{Introduction}

One of the main issues in
colossal magnetoresistance (CMR) manganites (La,$A$)MnO$_3$ is
to determine the appropriate model to understand the mechanism
of its magneto-transport phenomena.
Canonical model for these compounds is the  double-exchange (DE)
Hamiltonian introduced by Zener \cite{Zener51} in order to
study the ferromagnetism.
Transport properties in DE systems 
are recently studied in a controlled manner
using the 
dynamical mean-field approach by one of the authors \cite{Furukawa95all}.

Another candidate model for CMR manganites is
the system which incorporates large electron-lattice 
couplings \cite{Millis96l,Roder96}.
Effects of Anderson localization in the DE systems
with and without charge disorder \cite{Varma96,Allub96,Li96x} are also 
investigated.
Roles of orbital degeneracies in $e_{\rm g}$ electrons
have also been studied
in order to explain several transport properties  \cite{Ishihara96x,Shiba97}.

One way to justify the relevance of the model is to
compare universal behaviors insensitive to parametrization,
such as scaling relations, with experimental results.
For example, 
in compounds with relatively wide bandwidth $W$ and
high Curie temperature $T_{\rm c}$  such as (La,Sr)MnO$_3$ 
with sufficient doping concentration,
it has been shown that
magnetoresistance satisfy the scaling relation \cite{Tokura94}
\begin{equation}
  -\Delta\rho/\rho_0\propto   {M^*}^2
   \label{ScaleLSMO}
\end{equation}
 if not affected by extrinsic effects such as
grain boundary effects \cite{Hwang96}.
Here, $\rho_0$ is the resistivity at the paramagnetic phase
in the absence of magnetic field, $\Delta\rho$ is the
change in resistivity under magnetic field, and
$M^*$ is the magnetization normalized by its saturation value in the form
$M^* = M / M(T\to0)$. From theoretical point of view, this result 
is reproduced by the calculation within DE Hamiltonian
described above \cite{Furukawa95all},
which shows that this model is possibly a relevant model to
describe magneto-transport properties of these compounds 
at least in these scaling regime.

On the other hand, compounds with relatively 
narrow bandwidth $W$ and low $T_{\rm c}$ show
several different behaviors on magnetic and transport properties.
For example, scaling relation for resistivity and magnetism
which differs from eq.~(\ref{ScaleLSMO})
are reported for thin film data  in (La,Ca)MnO$_3$ \cite{Hundley95}.

Under the assumption of strong electron-lattice effect,
semiconductive behavior above $T_{\rm c}$ 
typically observed in these narrow $W$ compounds is explained by
formation of lattice polaron with dynamic JT distortion \cite{Millis96l}.
In these compounds,
large lattice distortions in oxygen are indeed observed
in (La,Ca)MnO$_3$ \cite{Radaelli95,Dai96,Billinge96}.
However, there are no reports so far which relates
theory with scaling relations in experiments for
low $T_{\rm c}$ compounds.
Therefore it is not clear at this moment whether lattice effect is
relevant or not to universalities in  magneto-transport properties
of narrow $W$ (low $T_{\rm c}$) compounds.


Investigation of spin excitations is another 
approach to study the nature of CMR manganites,
since the ferromagnetism in these systems reflects
the nature of electronic states \cite{Millis95}.
Experimentally, spin excitation spectrum of (La,$A$)MnO$_3$ ($A$=Pb, Sr, Ca)
has been measured by neutron inelastic scattering measurements
\cite{Martin96,Perring96,Hirota96,Lynn96,FernandezBaca97x}.
In La$_{0.7}$Pb$_{0.3}$MnO$_3$
which is classified as a wide $W$ compound with ferromagnetic
metal phase at low temperature, spin wave dispersion
in cosine-band form is observed throughout the Brillouin zone.

\begin{figure}
\epsfxsize=\HSIZE
\hfil\epsfbox{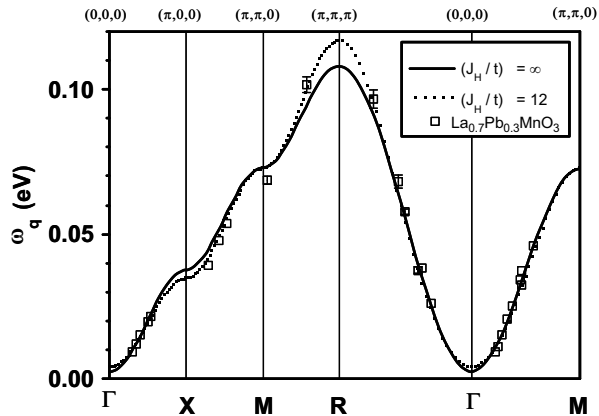}\hfil
\caption[]{Spin wave dispersion relation for the DE Hamiltonian
for $\JH/t=\infty$ (solid curve) and $\JH/t=12$ (dotted curve).
Experimental data (squares) are taken from ref. \cite{Perring96}.}
\label{FigSpinDispersion}
\end{figure}

Theoretically,
it is shown that double-exchange Hamiltonian on a cubic lattice reproduces
cosine-type dispersion relation
\begin{equation}
  \omega_q \propto 3- (\cos q_x + \cos q_y + \cos q_z)
\end{equation}
in the strong Hund's coupling limit
within the spin wave approximation, {\em i.e.} lowest order $1/S$ expansion.
In Fig.~\ref{FigSpinDispersion} we show the comparison between
the experiment \cite{Perring96} and theory  \cite{Furukawa96}.
We see that the DE system
also accounts for the spin wave excitation in CMR manganites.
From the fitting, we obtain the values for electron hopping
energy $t$ and Hund's coupling $\JH$. These parameters are also 
satisfactory in comparison
 with experiments in the way that $t$ is consistent
with the estimate for $e_{\rm g}$ electron bandwidth and
also in the way that the estimate for $T_{\rm c}$
for the DE model is consistent with experimental value \cite{Furukawa96}.

For the case of finite $S$, exact diagonalization studies have been
performed \cite{Zang96x,Kaplan97}. Quantum fluctuations at finite $S$
produces deviation from the cosine band, nevertheless the
deviation is small if doping concentration is
large enough \cite{Kaplan97}.
Therefore, as long as the dispersion relation is concerned,
DE Hamiltonian also accounts for the spin excitation of 
CMR manganites with wide $W$.

Anomalous temperature dependence in spin wave lifetime
is also  observed 
in La$_{0.7}$Pb$_{0.3}$MnO$_3$ \cite{Perring96}.
Line width of spin wave substantially
expanded by increasing temperature. 
This is typically observed at zone boundary,
{\em i.e.} high energy part of the spin wave dispersion relation.
Since the spin wave dissipation is determined by the 
presence of excitations coupled to spins,
study of spin wave lifetime clarifies not only the
spin excitation behavior but also the existence of excitation spectrum
for other degrees of freedom.

In this paper we investigate the lifetime of the spin wave
 excitations from both theoretical and experimental sides.
For theoretical approach, we study DE system, while experimentally
we study several compounds with relatively wide bandwidth.
We aim to obtain scaling relations in spin wave lifetime
and establish possible relevance between theory and experiment.

\section{Theoretical and Experimental Approaches and Results}

{}From the theoretical side, we study
the Kondo lattice model with
ferromagnetic spin exchange,
\begin{equation}
   \Ham = 
  - t \sum_{<ij>,\sigma}
        \left(  c_{i\sigma}^\dagger c_{j\sigma} + h.c. \right)
    -{J_{\rm H}}\sum_i 		\vec S_i \cdot \vec \sigma_i ,
\end{equation}
in the strong Hund's coupling region $\JH/t \gg 1$.
We consider the system on a cubic lattice with nearest neighbor
hoppings.
We investigate the spin wave lifetime in relation with
magnetic moment, and clarify the role of DE mechanism
with respect to dissipation of spin excitations.

Green's function $G_\sigma(k,\omega)
= (\omega-\varepsilon_k - \Sigma_\sigma(\omega))^{-1}$
at  finite temperature
is obtained from the dynamical mean-field approximation  \cite{Georges96}
on a cubic lattice.
Here, $\Sigma_\sigma(\omega)$ is a $k$-independent self energy.
Using these Green's functions,
Spin susceptibility $\chi(q,\omega)$
 are calculated through electron polarization
diagrams.
Within the lowest order of $1/S$ spin wave expansion \cite{Furukawa96},
real part of the self-energy for spin wave determines
the dispersion relation through the self-consistency equation
\begin{equation}
 \omega_q = \Re\Pi(q,\omega_q),
\end{equation}
and the self-energy is constructed from the spin polarization 
diagrams for fermions and vertices proportional to $J_{\rm H}$.
Spin wave line width $\Gamma_q$ is obtained by the imaginary part
of the self-energy,
\begin{equation}
 \Gamma_q = \Im\Pi(q,\omega_q) \propto \Im \chi(q,\omega_q).
\end{equation}
Physical interpretation of this result is that
the origin of the spin wave line width is the
dissipation process of spin wave into spin excitations in
 Stoner continuum.

\begin{figure}[htb]
\epsfxsize=\HSIZE
\hfil\epsfbox{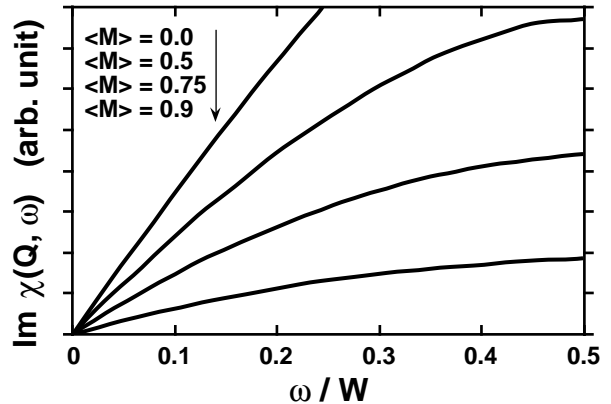}\hfil
\caption{Imaginary part of the Stoner susceptibility
(Stoner absorption) $\chi(Q,\omega)$ at the zone boundary
$Q=(\pi,\pi,\pi)$ for various temperatures.}
\label{FigImPi}
\end{figure}

\begin{figure}
\epsfxsize=\HSIZE
\hfil\epsfbox{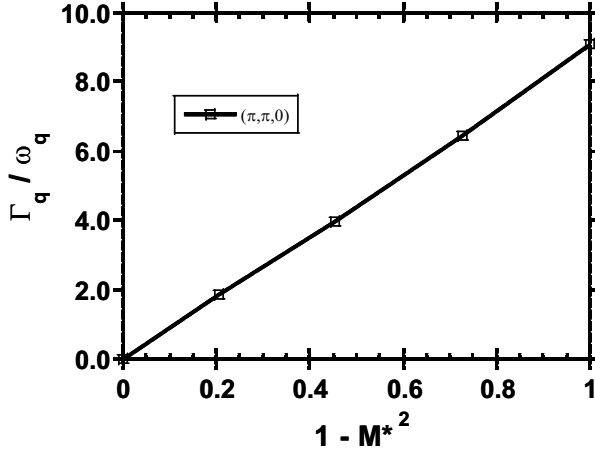}\hfil
\caption[]{Theatrical results for 
scaling behaviors in inverse lifetime of the spin wave in the form
$\Gamma_q / \omega_q \propto 1 - {M^*}^2$.}
\label{FigTheoryScale}
\end{figure}

In Fig.~\ref{FigImPi} we show ${\rm Im}\,\chi(Q,\omega)$
 at the Brillouin zone boundary $Q =
(\pi,\pi,\pi)$ for various temperatures, 
where we take $\JH/W=2$ and  $x=0.3$.
Here $W =6t$ is the electron bandwidth.
We see that at small $\omega$ we have $\omega$-linear
relation, {\em i.e.} 
$  \Im\chi \propto \omega$ at $\omega \ll W$.
Coefficients for $\omega$-linear part
  decrease by decreasing the temperature. 
In Fig.~\ref{FigTheoryScale} we show 
the coefficient for $\omega$-linear part 
$ \partial \Im\chi(Q,\omega)/ \partial \omega |_{\omega\to0}$
as a function of normalized magnetization $M^*$ at wave vector $Q$.
As a result we find
\begin{equation}
  \Im\chi(Q,\omega) \propto (1-{M^*}^2) \,\omega
   \label{ImChiScale}
\end{equation}
for small values of $\omega$.
The relation (\ref{ImChiScale})
is observed at all values of $q$ with slightly
different coefficients.
Since $\Gamma_q \propto \Im \chi(q,\omega_q)$, we
obtain the relation
\begin{equation}
  \Gamma_q = \alpha_q (1-{M^*}^2) \, \omega_q
	\label{ScalingFunction}
\end{equation}
where $\alpha_q$ is a dimensionless function weakly dependent on $q$.

Now, we show experimental results for spin wave lifetime
with respect to relatively wide $W$ (high $T_{\rm c}$) samples
of ((La,Y),Sr)MnO$_3$.  

Single crystals of (La$_{1-x-y}$Y$_{y}$)Sr$_{x}$MnO$_{3}$ (typically
$6\phi \times 80$~mm) were grown by the floating-zone method.  The end part
(30--40~mm) were used for neutron measurements.  In order to
study the effects of electronic band-width, three samples were chosen; $y=0.00,
0.05$ and 0.10 with the Sr concentration fixed at
$x=0.20$.  The tolerance factor $t$, which is defined as
$t=(r_{A}+r_{O})/\protect\sqrt{2}(r_{B}+r_{O})$ for ABO$_{3}$
perovskites and scales with the band width, decreases with
doping Y; 0.908, 0.906 and 0.903 for $y=0.00, 0.05$ and 0.10, respectively. 
These values are, however, still closer to unity than
La$_{0.8}$Ca$_{0.2}$MnO$_{3}$ (0.894) and Pr$_{0.8}$Ca$_{0.2}$MnO$_{3}$
(0.877). 
 Neutron-scattering measurements were carried out using the triple-axis
spectrometer TOPAN at the JRR-3M reactor in Japan Atomic Energy Research
Institute. Typical condition employed was the fixed final energy at 14.7~meV
with horizontal collimation of Blank-30$'$-Sample-60$'$-Blank.  The
$(002)$ reflection of pyrolytic graphite (PG) is used to monochromatize and
analyze neutrons.  A PG filter was used to reduce higher-order contaminations
in the incident beam.  The Curie temperature $T_{C}$ and its Gaussian
distribution $\Delta T_{C}$ was determined using the temperature dependence of
the $(100)_{Cubic}$ peak intensity; $T_{C}$ ($\Delta T_{C}$) is 306(1.1),
281(1.4) and 271(10)~K for $y=0.00, 0.05$ and 0.10, respectively.

\begin{figure}
\epsfxsize=\HSIZE
\hfil\epsfbox{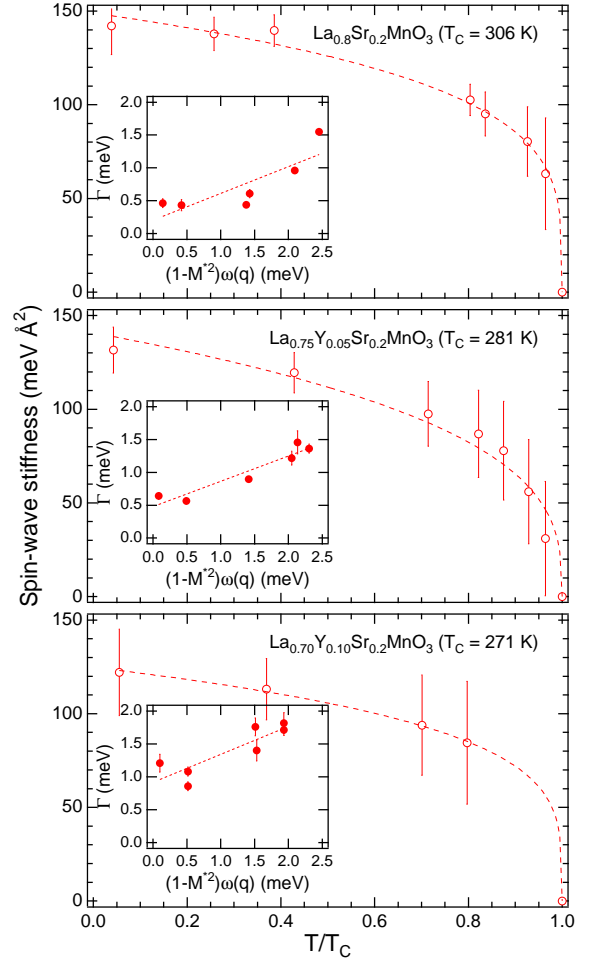}\hfil

\caption[]{Experimental results for the 
temperature dependence of the spin-wave stiffness $D$.  Error bars
indicate fitting errors.  Dashed lines are guides to the eye.  Inset shows a
linear relation between $\Gamma(q,T)$ and $(1-M^{*2})\omega(q,T)$, where
$M^{*}=M(T)/M(0)$.}
\label{Fig:Stiffness}
\end{figure}

The spin-wave dispersion curves for studied
(La$_{1-x-y}$Y$_{y}$)Sr$_{x}$MnO$_{3}$ samples show 
an isotropic behavior in the
measured ($q$, $\omega$) range,
$q<0.40$~\AA$^{-1}$ and $\hbar\omega<20$~meV, and follow well with
$\hbar\omega(q)=Dq^{2}+E_{0}$, typical of ferromagnetic spin-wave dispersion in
low $q$ and low $\omega$ range.  In order to study the temperature
dependence of spin-wave stiffness $D$, we have performed constant-$Q$
scans at (1.1\ 1.1\ 0) where well-defined peak profiles are obtained in a wide
temperature range. Each peak profile was fitted with the spin-wave
scattering cross section including the finite life-time $h/\Gamma$ convoluted
with a proper instrumental resolution. Figure \ref{Fig:Stiffness} shows thus
obtained temperature dependence of $D$.  Error bars
indicate fitting errors.  Dashed lines are guides to the eye.
Softening of spin wave dispersion $\omega(q,T)$ is observed
 as temperature approaches $T_{\rm c}$.

At the inset of Fig.~\ref{Fig:Stiffness} we show the inverse lifetime
$\Gamma(q,T)$ versus $(1-{M^*}^2)\, \omega(q,T)$.
From the figure we see that inverse lifetime is fitted in the form
\begin{equation}
  \Gamma(q,T) = \Gamma_0(q) + \alpha_q (1-{M^*}^2) \,\omega(q,T).
\end{equation}
Temperature dependence  is scaled in a way consistent with theoretical
result for DE systems. 
$\Gamma_0$ is the $T$- and $\omega$-independent part
of the spin wave lifetime, {\em i.e.} $\Gamma(q,T\to0) = \Gamma_0(q)$,
 which
increases  systematically by increasing the ratio of Y atoms.
We specurate that the origin of $\Gamma_0$ is mainly extrinsic effects 
such as disorder and inhomogeneity of the sample.

\section{Discussions and Summary}

\begin{figure}
\epsfxsize=\HSIZE
\hfil\epsfbox{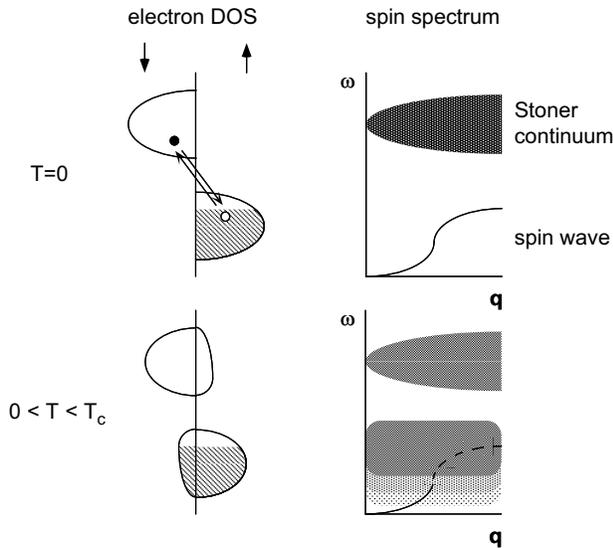}\hfil

\caption[]{Electron density of states and the Stoner excitation processes
of the DE systems at the ground state and at finite temperature.}
\label{FigStoner}
\end{figure}

This result is explained as follows.
At $T=0$ where local spin moment is saturated,
we have so called half-metallic state for itinerant electrons.
The Stoner excitation, which is a particle-hole process of
different species of spins, lies at high energy part
$\omega \sim 2\JH$. Then the spin wave excitation spectrum,
 can be
observed without being hidden by Stoner continuum.
This is the reason for observation of 
spin wave collective mode throughout the Brillouin zone 
 in the low 
temperature regime even though the system is metallic.
However, at finite temperature with unsaturated spin moment,
 the density of states splits
into two subbands \cite{Furukawa95all} 
and there appear low energy Stoner processes
which damp the spin wave modes.
In Fig.~\ref{FigStoner} we schematically illustrate
the Stoner processes at the ground state and at finite temperature.

Low energy part of the Stoner excitation is
produced by the process of occupied electron excited into
unoccupied state with opposite spin species. By counting
density of states available for the process
 we see $\Im\chi \propto \omega$ at $T\ll W$. 
As for temperature dependence, density of state change by temperature
in the way that the majority band has the weight $(1 + M^*)/2$ and
the minority band $(1 - M^*)/2$, 
which is the weight for the initial and final state of
Stoner processes, respectively.
 Thus the Stoner absorption is scaled in the form
$  \Im\chi \propto (1 - {M^*}^2) \, \omega$.

In the presence of strong electron-lattice coupling such as
Jahn-Teller mode, there should exist strong magnon damping
processes.
Presence of such strong damping is inconsistent with
the experimental fact for (La,Pb)MnO$_3$ that
spin wave excitation spectrum is well-defined throughout the Brillouin
zone \cite{Perring96}. Strong damping effect also
  cause the deviation from the scaling form (\ref{ScalingFunction}).
Thus, as long as spin wave excitation spectrum in relatively
wide band compounds are concerned, electron-lattice couplings
does not seem to be so strong as to affect magnon damping effects.

To summarize, we studied the spin wave lifetime
of the CMR systems, theoretically by dynamical mean-field approach
for the DE model and experimentally
by neutron inelastic scattering of ((La,Y),Sr)MnO$_3$.
Scaling relations in the spin wave
lifetime of DE model as well as CMR manganites is consistent
with each other.
Spin excitation spectrum of CMR manganites with
relatively wide bandwidth is explained by the DE model. 

\section*{Acknowledgement}
N.F. thanks G. Aeppli for suggestions and  discussions at the
early stage of this work, and  H.~Y. Hwang, A.~J. Millis and C. Varma for
useful comments.
K. H. acknowledges Y. Endoh for his valuable advises concerning the
experiments.
Numerical calculation is partially performed on the
 FACOM VPP500 at the Supercomputer Center, 
Inst.\ for Solid State Phys., Univ.\ of Tokyo.

%

\end{document}